\documentclass{PoS}

\def\dec{H^+\to W^+\gamma}

\def\dec{H^+\to W^+\gamma}
\def\PL{\it Phys. Lett.}

\def\PR{\it Phys. Rev.}

\def\beq{\begin{equation}}
\def\eeq{\end{equation}}
\def\[{\left[}
\def\]{\right]}

\def\gsim{\lower.7ex\hbox{$\;\stackrel{\textstyle>}{\sim}\;$}}
\def\lsim{\lower.7ex\hbox{$\;\stackrel{\textstyle<}{\sim}\;$}}

\title{Implications of Yukawa texture in the charged Higgs boson phenomenology within 2HDM-III}

\ShortTitle{ Yukawa texture in the charged Higgs boson}

\author{ A. Cordero-Cid \\
      Fac. de Cs. de la
Electr\'onica, Benem\'erita Universidad Aut\'onoma de Puebla, Apdo. Postal 542, Puebla, Pue. 72570, M\'exico.}

\author{ O. F\'elix-Beltr\'an \\
      Fac. de Cs. de la
Electr\'onica, Benem\'erita Universidad Aut\'onoma de Puebla, Apdo. Postal 542, Puebla, Pue. 72570, M\'exico.}

\author{\speaker{Jaime Hern\'andez-S\'anchez}\thanks{SNI-CONACYT and PROMEP-SEP grants}\\
       Fac. de Cs. de la
Electr\'onica, Benem\'erita Universidad Aut\'onoma de Puebla, Apdo. Postal 542, Puebla, Pue. 72570, M\'exico.\\
 Dual C-P Institute of High Energy Physics, Puebla, Pue., M\'exico.\\
        E-mail: \email{jaimeh@ece.buap.mx}}

\author{ R. Noriega-Papaqui\\
      Dual C-P Institute of High Energy Physics, Puebla, Pue., M\'exico.\\
\'Area Acad\'emica de Matem\'aticas y F\'{\i}sica,
Universidad Aut\'onoma del Estado de Hidalgo,
Carr. Pachuca-Tulancingo Km. 4.5, Pachuca, Hgo. 42184, M\'exico.}

\abstract{We discuss the implications of assuming a four-zero Yukawa texture for the properties of the charged Higgs boson within the context of the general 2-Higgs Doublet Model of Type III. We present  the charged Higgs boson couplings with heavy quarks and the resulting pattern for its decays, including the decay $H^+ \to W^+ \gamma$ at 1-loop level. The parameters chosen can still avoid the $B \to X_s \gamma$ constraint,
 the perturbativity and $\rho_0$ bound. Also, we present   the constraints of $ B0-\bar{B}0$ mixing and of the
radiative corrections to the  $Z  b \bar{b}$ vertex in the regime small $ \tan \beta$. The production of charged Higgs bosons is also sensitive to the modifications of its couplings, so that we also evaluate the resulting effects  on `direct' $c\bar{b}\to H^++c. c.$ and `indirect' $q\bar q,gg\to \bar t b H^++c. c.$ production. Significant scope exists at the Large Hadron Collider for several $H^\pm$ production and decay channels combined to enable one to distinguish between such a model and alternative 2-Higgs doublet scenarios.}

\FullConference{Third International Workshop on Prospects for Charged Higgs Discovery at Colliders - CHARGED2010,\\
		September 27-30, 2010\\
		Uppsala Sweden}

\begin{document}

\section{Introduction}

The 2HDM-II has been quite attractive to date, in part because it coincides
with the Higgs sector of the MSSM, wherein each Higgs doublet couples
to the $u$- or $d$-type fermions separately\footnote{Notice that there exist
significant
differences between the 2HDM-II and MSSM though, when it comes to their
mass/coupling configurations and possible Higgs signals \cite{Kanemura:2009mk}.}.
However, this is only valid at
tree-level \cite{Babu-Kolda}. Thus, we can consider the
2HDM-III as a generic description of physics at a higher scale (of
order TeV or maybe even higher), whose low energy imprints are
reflected in the Yukawa coupling structure. With this idea in
mind,  a detailed study of the 2HDM-III Yukawa
Lagrangian was presented in Refs.\cite{ourthdm3a, DiazCruz:2009ek}, under the assumption of a specific
texture pattern \cite{Fritzsch:2002ga}, which generalizes the
original model of Ref.~\cite{cheng-sher}. The extension of  such an approach
to investigate charged Higgs boson phenomenology was conducted in Ref.  \cite{DiazCruz:2009ek, BarradasGuevara:2010xs},  which discussed the implications of this Yukawa texture for the charged Higgs
boson properties (masses and couplings) and  the resulting pattern of
charged Higgs boson decays and main production reactions at the LHC.

This paper  is organized as follows. In section
2, we discuss the Higgs-Yukawa sector of the 2HDM-III. Then, in section
3, we shows the BR of the decays of the charged Higgs boson, including $\dec$  at one-loop level. Actual LHC
event rates for  the main
production mechanisms at the LHC are given in section 4. These
include the $s$-channel
production of charged Higgs bosons through
$c\bar{b}(\bar{c}b)$-fusion and the multi-body
more  $q \bar{q}, \, gg \to t \bar{b} H^- + $ c.c. (charge conjugated).  Finally, we summarize our results and present the conclusions in section 5. 

\section{The charged Higgs boson Lagrangian and the fermionic couplings}

 In order to derive the interactions of the charged Higgs boson, the
Yukawa Lagrangian is given by: 
\beq {\cal{L}}_{Y} =
Y^{u}_1\bar{Q}_L {\tilde \Phi_{1}} u_{R} +
                   Y^{u}_2 \bar{Q}_L {\tilde \Phi_{2}} u_{R} +
Y^{d}_1\bar{Q}_L \Phi_{1} d_{R} + Y^{d}_2 \bar{Q}_L\Phi_{2}d_{R},
\label{lagquarks} \eeq \noindent where $\Phi_{1,2}=(\phi^+_{1,2},
\phi^0_{1,2})^T$ refer to the two Higgs doublets, ${\tilde
\Phi_{1,2}}=i \sigma_{2}\Phi_{1,2}^* $, $Q_{L}$ denotes the
left-handed fermion doublet, $u_{R} $ and $d_{R}$ are the
right-handed fermions singlets and, finally, $Y_{1,2}^{u,d}$ denote the
$(3 \times 3)$ Yukawa matrices. Similarly, one can
write the corresponding Lagrangian for leptons.

After spontaneous EWSB  and including the
diagonalizing matrices for quarks and Higgs bosons\footnote{The
details of both diagonalizations are presented in
Ref.~\cite{ourthdm3a}.}, through rotated matrices $\tilde{Y}_n^{q} = V_q\, Y^{q}_n \, V_q^\dagger$ 
($n=1$ when $q=u$, and $n=2$ when
$q=d$ ) where $V_q$ is the diagonalizing mass matrix. 
One can derive a better approximation for the product
$V_q\, Y^{q}_n \, V_q^\dagger$, expressing the
rotated matrix $\tilde {Y}^q_n$, in the form
\begin{equation}
\left[ \tilde{Y}_n^{q} \right]_{ij}
= \frac{\sqrt{m^q_i m^q_j}}{v} \, \left[\tilde{\chi}_{n}^q \right]_{ij}
=\frac{\sqrt{m^q_i m^q_j}}{v}\,\left[\chi_{n}^q \right]_{ij}  \, e^{i \vartheta^q_{ij}},
\end{equation}
where $\chi$'s are unknown dimensionless parameters of the model, they come
from the election of a specific texture of the Yukawa matrices. We find 
the Lagrangian  of the interactions of the charged Higss boson with quark pairs as follows:
\begin{eqnarray}
\label{LCCH}
{\cal{L}}^{q} & = &
\frac{g}{2 \sqrt{2} M_W} \sum^3_{l=1} \bar{u}_i \left\{  (V_{\rm CKM})_{il} \left[ \tan \beta \, m_{d_{l}} \, \delta_{lj}
-\frac{\sec \beta}{\sqrt{2} }  \,\sqrt{m_{d_l} m_{d_j} } \, \tilde{\chi}^d_{lj}  \right] \right.
\nonumber \\
& & + \left[ \cot \beta \, m_{u_{i}} \, \delta_{il}
  -\frac{\csc \beta}{\sqrt{2} }  \,\sqrt{m_{u_i} m_{u_l} } \, \tilde{\chi}^u_{il} \right]
  (V_{\rm CKM})_{lj} \nonumber \\
& & + (V_{\rm CKM})_{il} \left[ \tan \beta \, m_{d_{l}} \, \delta_{lj}
-\frac{\sec \beta}{\sqrt{2} }  \,\sqrt{m_{d_l} m_{d_j} } \, \tilde{\chi}^d_{lj}   \right] \gamma^{5}
 \\
& & - \left. \left[ \cot \beta \, m_{u_{i}} \, \delta_{il}
  -\frac{\csc \beta}{\sqrt{2} }  \,\sqrt{m_{u_i} m_{u_l} } \, \tilde{\chi}^u_{il}  \right]
  (V_{\rm CKM})_{lj} \, \gamma^{5} \right\} \, d_{j} \, H^{+},   \nonumber
\end{eqnarray}
where we have redefined $\left[ \tilde{\chi}_{1}^u \right]_{ij} =
\tilde{\chi}^u_{ij}$ and $\left[ \tilde{\chi}_{2}^d \right]_{ij} =
\tilde{\chi}^d_{ij}$. Then, from Eq.~(\ref{LCCH}), the couplings $\bar{u}_i d_j H^+$ and $u_i \bar{d}_j
H^-$ are given by:
$ g_{H^+\bar{u_i}d_j} = -\frac{ig}{ 2\sqrt{2} M_W}
(S_{i j} +P_{i j} \gamma_5), \, g_{H^- u_i \bar{d_j}}= -\frac{ig
}{2\sqrt{2} M_W}  (S_{i j} -P_{i j} \gamma_5)$, where
$S_{i j}$ and $P_{i j}$ are defined as:
$
^{S_{i j}}_{P_{ij}} =   \sum^3_{l=1} (V_{\rm CKM})_{il}  \, m_{d_{l}} \, X_{lj}
 \pm  m_{u_{i}} \, Y_{il}  (V_{\rm CKM})_{lj}.
$
with
$
X_{l j}  =    \bigg[ \tan \beta \,
\delta_{lj} -\frac{\sec \beta}{\sqrt{2} }  \,\sqrt{\frac{m_{d_j}}{m_{d_l}}  }
\, \tilde{\chi}^d_{lj}  \bigg],
\, \, \,
Y_{i l} =   \bigg[ \cot \beta \,  \delta_{il}
  -\frac{\csc \beta}{\sqrt{2} }  \,\sqrt{\frac{ m_{u_l}}{m_{u_i}} } \, \tilde{\chi}^u_{il}  \bigg] .
$
Based on the analysis of $B \to X_s
\gamma$ \cite{Borzumati:1998nx, Xiao:2003ya}, it is claimed that $X
\leq 20$ and $Y \leq 1.7$ for $m_{H^+} > 250$ GeV, while for a
lighter charged Higgs boson mass, $m_{H^+} \sim 180$ GeV, one gets
$(X,Y) \leq (18,0.5)$. Thus, we find
the bounds: $|\chi_{33}^{u,d}| \lsim 1$ for $0.1 < \tan \beta
\leq 70$ \cite{DiazCruz:2009ek}. 
On the other hand, the condition $\frac{\Gamma_{H^+}}{m_{H^+}} < \frac{1}{2}$ in the
frame of  the 2HDM-III implies 
$\frac{\Gamma_{H^+}}{m_{H^+}} \approx \frac{3G_F
m_t^2}{4\sqrt{2}\pi\tan\beta^2} \bigg(
\frac{1}{1-\frac{\tilde{\chi}^u_{33}}{\sqrt{2} \cos\beta}}\bigg)^2$,
we have checked numerically that this leads to $0.08 < \tan\beta <
200$ when $|\tilde{\chi}^u_{33}| \approx 1$ and  $0.3 < \tan\beta <
130$ as long as $|\tilde{\chi}^u_{33}| \to 0$ recovering the result
for the case of the 2HDM-II \cite{Chankowski:1999ta}.

Another important bounds on $\vert \tilde{\chi}_{33}\vert$  and tan$\beta$
comes from  radiative corrections to the  process $\Gamma(Z \to b \bar{b})$, specially the hadronic branching fraction of 
$Z$ bosons to $b\bar{b}$ ($R_b$) and the $b$ quark asymmetry ($A_b$) impossed a high restriction.
Then, following the calculation of the Ref. \cite{BarradasGuevara:2010xs}, in the regime of small $\tan\beta $, we can find bounds for  $\tan \beta$:  in the case $\chi_{33}^{u,d} = 1 $ and $m_{H^+} \sim 200 (300)$ GeV,  the  range
$\tan \beta > 0.3 (0.2)$ is allowed, while in the scenario $\chi_{33}^{u,d} = -1 $ and $m_{H^+} \sim 200 (300)$ GeV, 
$\tan \beta > 5 (3)$ is permitted. 

In the
Ref. \cite{BarradasGuevara:2010xs} is presented the analysis of the quantity that parameterizes the $B_0-\bar{B_0}$ mixing: $x_d \equiv \frac{\Delta m_B}{\Gamma_B} $, where we obtain bounds for $\tan \beta$ and mass of the charged Higgs boson.
Combining the criteria of the analysis radiative corrections of $Z b \bar{b}$ vertex and  $ B_0-\bar{B}_0$ mixing,  $\tan \beta > 0.3$ is allowed for 
$m_{H^+} > 170 $ GeV and $ \chi_{33}^{u,d}=1$. However, when  $ \chi_{33}^{u,d}=-1$  and  $m_{H^+} <600 $ GeV, $\tan \beta < 2$ is disfavored.

Besides, following the analysis of the Ref. \cite{BarradasGuevara:2010xs}, one can get the deviation $\Delta \rho_0$ of the parameter 
$\rho_0= M_W^2/ \rho M_Z^2 C^2_W$ of our version 2HDM-III, where the $\rho$ in the denominator absorbs all the SM corrections, among which
the most important SM correction at 1-loop level comes from the heavy
top-quark.  We can get that for the case $\alpha = 0, \pi/2$, the parameter space of the scalar sector is strongly reduced when  decoupling between Higgs bosons, i.e. 
$\Delta m_{ij}= m_i-m_j > 100$ GeV ($m_i= m_{h^0}, \,m_{H^0}, \, m_{A^0}, \,m_{H^\pm}$). However, is possible to avoid the constraint for $\Delta \rho_{\rm 2HDM-III} $ if  the decoupling  source $\Delta m_{ij} \sim 20$ GeV o $\Delta m_{ij} \sim 100$ GeV but one Higgs very heavy (e.g. $m_{H^0}> 1$ TeV). When 
 $\alpha = \beta \pm \pi/2$ the allowed parameter region is larger and one can avoid the constraints of the $\rho$ parameter with or without decoupling. 

\section{Decays of the charged Higgs boson at tree level}

Let us now discuss the decay modes of the charged Higgs boson
within our model. Hereafter, we shall refer to scenario with
$\tilde{\chi}^u_{ij}=1$, $\tilde{\chi}^d_{ij}=1$ and $\tan\beta=0.3, \, 0.5, \, 1,
\, 10$. We have performed the numerical analysis
of charged Higgs boson decays, taking the mixing angle $\alpha =  \pi/2+\beta$ and varying the charged Higgs boson mass within the interval
100 GeV $\leq m_{H^{+}} \leq$ 800 GeV, further fixing $m_{h^0}= 120$
GeV, $m_{A^0}\sim m_{H^+}$ \cite{BarradasGuevara:2010xs}.

\begin{figure}
\centering
\includegraphics[width=6in]{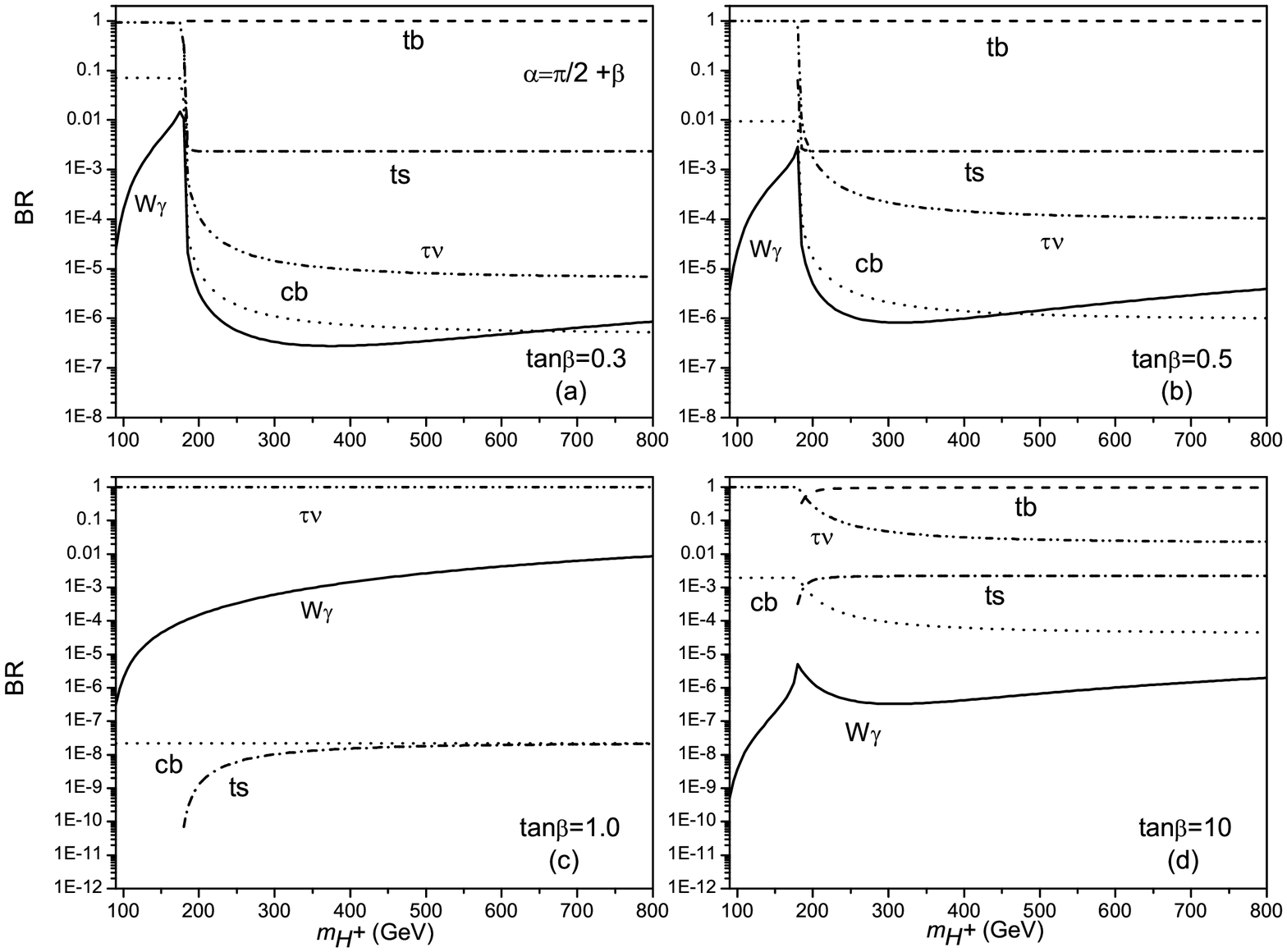}
\caption{The figure shows the BRs of the $H^+$ decaying
into the principal modes, taking
$\tilde{\chi}_{ij}^{u}=1$, $\tilde{\chi}_{ij}^{d}=1$, $m_{h^0} =
120$ GeV, $m_{A^0}= m_{H^+}$ and $\alpha = \pi/2+ \beta$  for: (a) $\tan
\beta=0.3$, (b) $\tan \beta = 0.5$, (c) $\tan \beta = 1$, (d) $\tan
\beta = 10$. The lines  in each graph correspond to: ($W \gamma$ line) BR($\dec $), ($tb$ line) BR($H^+ \to
t\bar{b} $), ($cb$ line) BR($H^+ \to c\bar{b}$), ($ts $ line) BR($H^+ \to t\bar{s}$),
  ($\tau \nu$ line) BR($H^+ \to \tau^+ \nu_\tau$), (Wh line) BR($H^+ \to W^+ h^0$).} \label{fig:wga-2a}
\end{figure}
 We present this special and interesting case  in the Fig. \ref{fig:wga-2a}. 
For $\tan \beta = 0.3$, we show  in plot Fig.~\ref{fig:wga-2a}(a) that  the relevant channel decay is the $\tau^+ {\nu_\tau} $ when $m_{H^+}<180 $ GeV, and for the range
 $m_{H^+}>$ 180 GeV the mode $t \bar{b}$ becomes dominant. Here, the BR of the 
 decay $\dec$ is relatively large of order $10^{-2}$. From Fig.  \ref{fig:wga-2a} (b) we can observe that the dominant decay mode is into $\tau^+ {\nu_\tau}$ for the
range $m_{H^+}<$ 175 GeV, again for 175 GeV $< m_{H^+} < 180$ GeV
the mode $t\bar{s}$ is the leading one, while the mode $W^+ \gamma$ induced at one-loop level has a BR  of order $10 ^{-3}$. When 
 $m_{H^+}>$ 180 GeV the mode $t \bar{b}$ is the leading one. The most interesting case is when $\tan \beta =1$ where the width decay of the mode $t \bar{b}$ is zero. We present this case in the  Fig. \ref{fig:wga-2a} (c), where one can see that the dominant decay is the mode $\tau^+ \nu_\tau$ for all  $m_{H^+}$. Besides, the BR $(\dec) $ is of order $10^{-2}$ to  $10^{-4}$  for   $m_{H^+}>$ 180 GeV. Now, from
Fig.~\ref{fig:wga-2a}(c), where $\tan \beta = 10$, we find that
 the dominant decay mode is into $\tau^+ {\nu_\tau}$ for the
range $m_{H^+}<$ 180 GeV. For
180 GeV$<m_{H^+}$, the dominant decay of the charged Higgs
boson is the mode  $t \bar{b}$.
We observe that the mode $W^+ \gamma$
is important when 170 GeV $<m_{H^+} <$ 180 GeV and for $0.1 \leq
\tan \beta \leq 1$, taking $\tilde{\chi}_{ij}^{u,d}=1$.  

\section{Event rates of charged Higgs bosons at the LHC}

To illustrate the type of charged Higgs signatures that have the
potential to be detectable at the LHC in the 2HDM-III, we show in
Tabs.~\ref{tab:5} and \ref{tab:6}  the event rates of charged Higgs
boson through the channels $q\bar q,gg\to t\bar b H^-_i$ + c.c. and
$c\bar{b} \to H^+$ + c.c., alongside the corresponding production
cross sections ($\sigma$'s) and relevant BRs, for a combination of
masses, $\tan\beta$ and specific 2HDM-III parameters amongst those
used in the previous works \cite{DiazCruz:2009ek, BarradasGuevara:2010xs} (assuming $m_{h^0}= 120$ GeV,
$m_{A^0}=300$ GeV and the mixing angle at $\alpha = \pi /2$
throughout). In particular, we focus on those cases where the
charged Higgs boson mass is above the threshold for $t \to b H^+$. (As default, we also assume an
integrated luminosity of $10^5$ pb$^{-1}$.)

To illustrate these results, let us comment on one case within each
scenario. From Table \ref{tab:5}, we can see that for Scenario
with $(\tilde{\chi}_{ij}^{u}=1, \tilde{\chi}_{ij}^{d}=1)$ and
$\tan\beta=15$, we have that the ${H^{\pm}}$ is heavier than $m_t -m_b$,
as we take a mass $m_{H^+}=400$, thus precluding top decay
contributions, so that in this case $\sigma(pp \to t \bar{b} H^+)
\approx 2.2 \times 10^{-1}$ pb, while the dominant decays are $H^+
\to t \bar{b}, \tau^+ \nu_\tau \, W^+ h^0,  \, W^+ A^0$ which give
a number of events of 7040, 46, 13860, 374, respectively. In this
case the most promising signal is $H^+ \to W^+ h^0$. However, when
$\tan \beta=70$ we have that all event rates increase substantially. Here,
the signal $H^+ \to W^+ h^0$ is still the most important with an
event rate of 15480.

All these rates correspond to the case of indirect production. The
contribution due to direct production is in fact subleading,
especially at large $m_{H^\pm}$ values. Nonetheless, in some
benchmark cases, they could represent a sizable addition to the
signal event rates. This is especially the case for 
$\tan\beta=15$ or 70. In general
though, also considering the absence of an accompanying trigger
alongside the $H^\pm$,{\it i.e.} for instance a top quark produced
in $g b \to H^- t$ could help to identify the signal. Thus, we
expect that the impact of $c\bar b$-fusion at the LHC will be more
marginal that that of $gg$-fusion for large Higgs masses, in fact,
at times even smaller that the contribution from $q\bar
q$-annihilation \cite{DiazCruz:2009ek}.

\begin{table*}[htdp]
\begin{center}
\caption{\label{tab:5} Summary of LHC event rates for some parameter combinations with 
 $(\tilde{\chi}_{ij}^{u}=1, \tilde{\chi}_{ij}^{d}=1)$ and integrated luminosity of $10^{5}$
pb$^{-1}$, for several different signatures, through the channel $q
\bar q,gg \to \bar{t} b H^+$ + c.c.}
{\footnotesize
\begin{tabular}{|c|c|c|c|c|c|}
\hline $(\tilde{\chi}_{ij}^{u}, \tilde{\chi}_{ij}^{d})$  & $\tan\beta$ & $m_{H^+}$ (GeV) &
$\sigma(pp \to H^+ \bar{t} b)$ (pb) & Relevant BRs & Nr. Events\\

\hline \multicolumn{1}{|c|}{ (1,1)  }& 15 &400 &
$2.23\times 10^{-1}$ &\begin{tabular}{l} BR$\left( H^{+} \to
t\bar{b}\right)\approx 3.2 \times 10^{-1} $\\
BR$\left( H^{+} \to
\tau^{+}\nu^{0}_\tau\right) \approx 2.1 \times 10^{-3} $\\
BR$\left( H^{+} \to
W^{+}h^{0}\right) \approx 6.3 \times 10^{-1} $\\
BR$\left( H^{+}_{2} \to
W^{+}A^{0}\right) \approx 1.7 \times 10^{-2} $%
\end{tabular}
& \multicolumn{1}{|c|}{$
\begin{tabular}{r}
7040\\
46\\
13860\\
374%
\end{tabular}
$} \\ \hline

\hline \multicolumn{1}{|c|}{ (1,1) }& 70 & 400 &
$4.3\times 10^{-1}$ &\begin{tabular}{l} BR$\left( H^{+} \to
t\bar{b}\right)\approx 3.5 \times 10^{-1} $\\
BR$\left( H^{+} \to
c \bar{b}\right) \approx 1.4 \times 10^{-2} $\\
BR$\left( H^{+} \to
\tau^{+}\nu_{\tau}\right) \approx 2.5 \times 10^{-1} $\\
BR$\left( H^{+} \to
W^{+}h^{0} \right) \approx 3.6 \times 10^{-1} $%
\end{tabular}
& \multicolumn{1}{|c|}{$
\begin{tabular}{r}
15050\\
602\\
10750\\
15480%
\end{tabular}
$} \\ \hline

\end{tabular} }
\label{default5}
\end{center}
\end{table*}
\begin{table*}[htdp]
\begin{center}
\caption{\label{tab:6} Summary of LHC event rates for some parameter combinations with 
$(\tilde{\chi}_{ij}^{u}=1, \tilde{\chi}_{ij}^{d}=1)$
and integrated luminosity of $10^{5}$
pb$^{-1}$, for several different signatures, through the channel $c
\bar b \to H^+$ + c.c.}
{\footnotesize
\begin{tabular}{|c|c|c|c|c|c|}
\hline $(\tilde{\chi}_{ij}^{u}, \tilde{\chi}_{ij}^{d})$  & $\tan\beta$ & $m_{H^+}$ (GeV) &
$\sigma(pp \to H^+ + X)$ (pb) & Relevant BRs & Nr. Events\\

\hline \multicolumn{1}{|c|}{ (1,1) }& 15 &400 & $ 1.14 \times
10^{-1}$ &\begin{tabular}{l}  BR$\left( H^{+} \to
t\bar{b}\right)\approx 3.2 \times 10^{-1} $\\
BR$\left( H^{+} \to
\tau^{+}\nu^{0}_\tau\right) \approx 2.1 \times 10^{-3} $\\
BR$\left( H^{+} \to
W^{+}h^{0}\right) \approx 6.3 \times 10^{-1} $\\
BR$\left( H^{+}_{2} \to
W^{+}A^{0}\right) \approx 1.7 \times 10^{-2} $%
\end{tabular}
& \multicolumn{1}{|c|}{$
\begin{tabular}{r}
3648\\
24\\
7182\\
194%
\end{tabular}
$} \\ \hline

\hline \multicolumn{1}{|c|}{ (1,1) }& 70 & 400 & $1.25\times
10^{-1}$ &\begin{tabular}{l} BR$\left( H^{+} \to
t\bar{b}\right)\approx 3.5 \times 10^{-1} $\\
BR$\left( H^{+} \to
c \bar{b}\right) \approx 1.4 \times 10^{-2} $\\
BR$\left( H^{+} \to
\tau^{+}\nu_{\tau}\right) \approx 2.5 \times 10^{-1} $\\
BR$\left( H^{+} \to
W^{+}h^{0} \right) \approx 3.6 \times 10^{-1} $%
\end{tabular}
& \multicolumn{1}{|c|}{$
\begin{tabular}{r}
4375\\
175\\
3125\\
4500%
\end{tabular}
$} \\ \hline

\end{tabular} }
\label{default6}
\end{center}
\end{table*}

\section{Conclusions}

We have discussed the implications of assuming a four-zero Yukawa
texture for the properties of the charged Higgs boson, within the
context of a 2HDM-III. The latter clearly
reflect the different coupling structure of the 2HDM-III, e.g.,
with respect to the 2HDM-II, so that one has at disposal more
possibilities to search for $H^{\pm}$ states at current and future
colliders, ideally enabling one to distinguish between different
Higgs models of EWSB. We have then concentrated our analysis to
the case of the LHC and showed that the production rates of
charged Higgs bosons at the LHC is sensitive to the modifications
of the Higgs boson couplings. Finally, we have determined the number of events
for the most promising LHC signatures of a $H^\pm$ belonging to a
2HDM-III, for both $c \bar{b}\to H^+$ + c.c. and $q\bar q\to \bar
t bH^+$ + c.c. scatterings (the latter affording larger rates than
the former). Armed with these results, we are now in a position to
carry out a detailed study of signal and background rates, in
order to determine the precise detectability level of each
signature. However, this is beyond the scope of present work and
will be the subject of a future publication.

\end{document}